\documentclass[aps,twocolumn,prl]{revtex4}

\pdfoutput=1

\usepackage[utf8]{inputenc}
\usepackage{lmodern}
\usepackage{hyperref}
\usepackage{amsmath}
\usepackage{amssymb}
\usepackage{graphicx}
\usepackage{xcolor}
\usepackage{color}
\usepackage{amssymb}
\usepackage{amsthm}
\usepackage{epstopdf} 
\usepackage{natbib}
\usepackage{dcolumn}
\usepackage{amsfonts}
\usepackage{bm}

\newcommand{\cc}{\,\mathrm{cm^{-3}}}
\newcommand{\wcm}{\,\mathrm{W/cm}^2}
\newcommand{\mic}{\,\mu\mathrm{m}}
\newcommand{\nm}{\,\mathrm{nm}}
\newcommand{\mev}{\,\mathrm{MeV}}

\newcommand{\fs}{\,\mathrm{fs}}

\newcommand{\rad}{\,\mathrm{rad}}
\newcommand{\mrad}{\,\mathrm{mrad}}

\newcommand{\micron}{\,\mu\mathrm{m}}

\begin{document}

\title{Relativistic acceleration of electrons injected by a plasma mirror into a radially polarized laser beam}

\author{N. Zaïm, M. Thévenet, A. Lifschitz, J. Faure}
\address{LOA, ENSTA ParisTech, CNRS, Ecole polytechnique, Universit\'e Paris-Saclay, 828 bd des Mar\'echaux, 91762 Palaiseau cedex France}

\begin{abstract}

We propose a method to generate femtosecond, relativistic and high-charge electron bunches using few-cycle and tightly focused radially polarized laser pulses. In this scheme, the incident laser pulse reflects off an overdense plasma that injects electrons into the reflected pulse. Particle-In-Cell simulations show that the plasma injects electrons ideally, resulting in a dramatic increase of charge and energy of the accelerated electron bunch in comparison to previous methods. This method can be used to generate femtosecond pC bunches with energies in the 1-10 MeV range using realistic laser parameters corresponding to current kHz laser systems.

\end{abstract}

\maketitle

The advent of femtosecond lasers that can reach enormous intensities, well beyond $10^{18} \wcm$~\cite{yu12}, brings about new possibilities. Among these, the acceleration of electrons to relativistic energies in very short distances is particularly promising. One of the many advantages of using lasers to accelerate electrons is the possibility to create ultrashort relativistic electron beams, with durations of a few femtoseconds~\cite{lund11}. As of today, laser wakefield accelerators \cite{esar09} have paved the way to electron acceleration in the 100~MeV  to multi-GeV range using 100~TW to PW laser drivers \cite{lund11,leem14}, with low repetition rates. However, there is also a need for electron sources with MeV energy and a high-repetition rate. Indeed, applications such as ultrafast imaging and diffraction and femtosecond pulse radiolysis require lower energy electrons but high reliability and statistics, calling for high-repetition rate operation~\cite{zewa06,scia11,koza00}. The aformentioned GeV beams are inappropriate for these applications both because of their low repetition rate and the small diffusion cross sections at these high energies. Despite recent progress of kHz laser wakefield accelerators in the MeV range~\cite{He13,beau15,salehi17,guen17}, Vacuum Laser Acceleration (VLA) is a very good candidate for producing MeV electrons at high-repetition rates. Indeed, in VLA, electrons are directly accelerated using the laser fields and the energy gain scales as $\Delta W[\mev]=30 \sqrt{P[TW]}$ ~\cite{esar95}, showing that MeV acceleration is possible with sub-TW laser systems. This is of great interest because TW peak powers have been demonstrated in milliJoule and kiloHertz laser systems, by post-compressing laser pulses to few cycle, sub-5-fs duration~\cite{bohm10,boeh14}. VLA using such lasers could enable the development of kHz femtosecond electron sources.

VLA with radially polarized laser beams (rather than with linearly polarized beams) was proposed as a way to improve the electron beam quality~\cite{esar95,sala06,karm07,wong10,paye12,vari13,marc13b,marc13,marc15,carb16}. Indeed, such beams contain both features of an efficient accelerator: an accelerating field in the longitudinal direction $E_z$ as well as a radial field $E_r$ that can confine electrons close to the optical axis. The longitudinal field $E_z$ becomes more significant as the beam is tightly focused. Numerical simulations have shown that VLA with radial polarization resulted in more energetic and more collimated electron beams than VLA with linear polarization~\cite{karm07}. Two schemes are commonly proposed for injecting and accelerating electrons into radially polarized laser pulses: (i) ionization injection, where electrons are released near a maximum of the laser electric field by ionizing a low density gas~\cite{marc13} and (ii) external beam injection, where a pre-accelerated electron beam is injected into the laser fields~\cite{carb16}. At present only two experimental results have been published, in which energy gains ranging from a few keV to tens of keV have been obtained~\cite{paye12,carb16}. In this letter, we show that these modest energy gains are due to non-optimal injection conditions. We propose a simple method based on the use of an overdense plasma for optimally injecting and accelerating electrons into radially polarized laser pulses. We show that high-charge (several pC), relativistic (1-10 MeV) beams with femtosecond durations can be obtained with current mJ lasers operating at a kHz repetition rate.

We start by explaining how the initial injection condition affect the subsequent acceleration of electrons. For simplicity, we first consider the linear polarization case, with a laser electric field $E_L\propto \sin\phi$, where $\phi=\omega_0t-k_0z$ is the electron phase in a field with wave vector $k_0$ and angular frequency $\omega_0$. In a plane wave, it is straightforward to show~\cite{hart95} that the maximum energy gain is $\Delta W_{max} = m_ec^2\gamma_0a_0^2(1+|\cos\phi_i| )^2$. Here, $\gamma_0$ is the initial Lorentz factor of the electron, $a_0$ is the normalized amplitude of the field, given by $a_0 = E_{L,max}/E_0$ with $E_0 = m_ec\omega_0/e$, and $\phi_i$ is the initial phase of the electron in the laser. This formula shows that the energy gain is highest when: (i) using high amplitude laser fields, (ii) injecting electrons inside the laser pulse at the optimal phase, $\phi_i=0$, \textit{i.e.} at a zero of the electric field, and (iii) injecting electrons with a high initial energy $\gamma_0$. These specific initial conditions are difficult to achieve experimentally, explaining why the observation of VLA has been a challenge. For radially polarized laser pulses, the problem is intrinsically 3-dimensional and obtaining simple analytical estimates is quite difficult. Therefore, in the following, we use numerical calculations to show that the optimum injection conditions are the same as in the linear polarization case. 

Within the framework of the paraxial approximation and assuming a Gaussian temporal envelope, the on-axis longitudinal electric field of the lowest order radially polarized laser beam is~\cite{sala06} :

\begin{equation}
E_z(z,t) = E_{z0} \frac{w_0}{w(z)} \sin\phi \, \exp\left(-2\log(2)\frac{\left(t-\frac{z}{c}\right)^2}{\tau_0^2}\right)
\label{paraxial}
\end{equation}

With $\phi = \omega_0t-k_0z+2\arctan(z/z_0)-\phi_0$. $w(z) = w_0\sqrt{1+\frac{z^2}{z_0^2}}$ is the beam waist, $w_0$ the minimum beam waist, $z_0 = k_0w_0^2/2$ is the Rayleigh length, $\tau_0$ is the pulse duration in FWHM of the intensity and $\phi_0$ is the initial phase. We also define $a_{0z}$ as the normalized amplitude of the longitudinal field : $a_{0z} = E_{z0}/E_0$ with $E_0$ defined earlier. Because of the Gouy phase $2\arctan(z/z_0)$, the axial phase velocity of the beam is superluminous~\cite{vari13}. If the interaction between the electron and the laser is limited between $z = 0$ (\textit{i.e.} the beam waist) and $z = +\infty$, as is the case for most proposed accelerating schemes, the phase difference due to the Gouy phase is $\pi$. As a consequence, electrons cannot stay indefinitely in an accelerating phase of the laser (where $E_z$ is negative). An electron can reach high energies if it remains in an accelerating half-cycle for a long time and net energy gains can be obtained if the subsequent decelerating half-cycle is diminished due to diffraction or the temporal shape of the pulse. An electron can stay in an accelerating phase longer if it has a velocity close to the speed of light, explaining why it is advantageous to inject electrons with a high initial energy or to use a high laser amplitudes.

\begin{figure}
\includegraphics[width=\columnwidth]{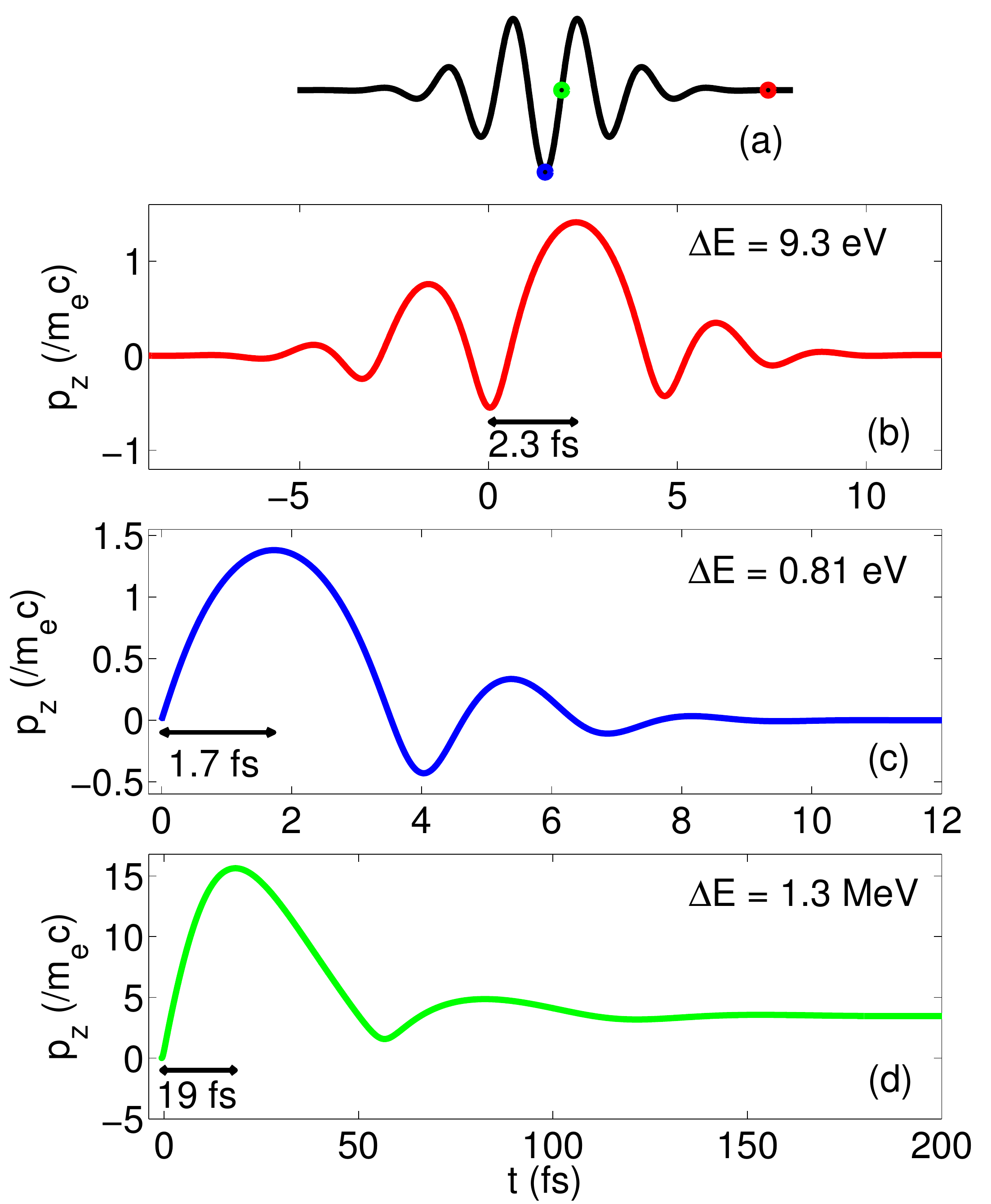} 
\caption{Results from test particle simulations. (a) Waveform of the few-cycle longitudinal electric field. The colored dots represent the initial positions of the electron in (b)-(d). (b)-(d) Longitudinal momentum $p_z$ along the electron trajectory for various cases: the electron is initially at rest either (b) in front of the pulse, (c) inside the pulse at a maximum of the field, (d) inside the pulse at a zero of the field. The double headed arrows show the time spent inside the main accelerating half-cycle.}
\label{singlep}
\end{figure}

The efficiency of the acceleration can also by greatly improved by carefully choosing the initial phase of the electron. To illustrate this, we perform on-axis test particle simulations of an electron initially at rest at $r=z=0$ that is accelerated by the field in Eq.~\ref{paraxial}, for three different initial phases. To model current kHz laser systems~\cite{boeh14}, we use the following parameters: $\lambda_0 = 800 \nm$, $a_{0z} = 0.7$, $w_0 = 1.5 \mic $, $\tau_0 = 3.5 \fs$ and $\phi_0 = \pi/2$. With these values, using the paraxial approximation and a Gaussian envelope is not perfectly valid but can nonetheless lead to decent estimates for the on-axis energy gain~\cite{marc13b}. Figures~\ref{singlep}(b) and~\ref{singlep}(c) show trajectories for non optimal initial phases, where the electron starts respectively in front of the laser pulse and inside the pulse at a maximum of the electric field. This is similar to the case of the ionization of a gas with respectively a low ionization energy and a high ionization energy. In each of these cases, the electron quickly dephases, resulting in negligible energy gains (respectively 9.3 eV and 0.81 eV). On the opposite, Fig.~\ref{singlep}(d) shows more efficient acceleration when the electron is initially at a zero of the electric field. This is because the electron starts with a whole accelerating half-cycle and stays in it much longer than in the previous cases, resulting in a final energy of 1.3 MeV. The above considerations show that VLA with radial polarization can lead to strong acceleration if electrons are injected (i) inside the laser pulse (ii) at a zero of the longitudinal field and (iii) with a high velocity in the propagation direction. While no current experimental setup allows such sub-wavelength precision, we show that these condition can be naturally satisfied using a plasma mirror injector.

A plasma mirror is an overdense plasma with a short density gradient, with a typical scale length of $<\lambda_0/10$. When an ultraintense laser pulse is focused on such an overdense plasma, it reflects off the plasma which behaves like a nearly perfect mirror (hence the term \emph{plasma mirror}). Nevertheless, while the laser pulse interacts with the plasma density gradient, it is able to pull out electrons and inject them into the reflected pulse with ideal initial conditions, allowing them to be efficiently accelerated. Recent experiments using gaussian lasers with linear polarization led to a clear observation of electron acceleration to energies in the MeV range~\cite{thev15}, but with rather large divergence angles of tens of degrees~\cite{thev15,bocoum16}. We demonstrate in the following that the concept of plasma mirror injection can also be applied to radially polarized laser pulses, potentially leading to more efficient acceleration and better beam quality. 
 
We used PIC simulations with the quasi-3D code CALDER-Circ~\cite{lifs09} to confirm that plasma mirrors fittingly allow us to inject a highly charged bunch of electrons near the zero of the electric field and with an initial velocity of a few hundreds of keV. Thanks to these optimal initial conditions, the electrons are then accelerated to relativistic energies by the reflected pulse. In our simulation, the plasma has a maximum electron density of $200 \, n_c$, with $n_c = 1.7 \times 10^{21} \cc$. The density decays exponentially with a gradient length of $\lambda_0/7$. The laser beam is focused on the plasma at normal incidence, and a moving window is started after the interaction, making it possible to follow the ejected electrons far from the plasma. Following~\cite{marc13}, we use an exact closed-form solution with a Poisson-like spectrum to model the radially polarized pulse (more details can be found in the Supplemental Material section). We take the same laser parameters as for the single particle simulations: $\lambda_0 = 800 \nm$, $a_{0z} = 0.7$, $\phi_0 = \pi/2$, $w_0 = 1.5 \mic $ and $\tau_0 = 3.5 \fs$. Numerical parameters can be found in the Supplemental Material.

Figure~\ref{video} displays three different snapshots from the PIC simulation, showing the interaction between the laser and the plasma. The electrons are ejected via the push-pull mechanism that was identified and fully described for linearly polarized beams in~\cite{thev16}. It consists of the following two steps. (1): The normal component of the electric field of the laser $E_z$ pushes electrons inside the plasma, resulting in a density peak. As the electron density peak is pushed deeper into the density gradient, the immobile ions create a large restoring static field, resembling a plasma capacitor. (2): When the density peak reaches its maximum depth  (Fig.~\ref{video}(b)), the sign of the electric field switches and both the laser and the static field work together to pull the electrons out of the plasma. A small fraction of the electrons inside the density peak can gain enough energy from the plasma capacitor to be ejected from the plasma. These electrons are ideally injected into the reflected pulse since they start with an initial velocity at the optimal phase, where the sign of the longitudinal field changes, and thus start with an accelerating half cycle, represented in blue in Fig.~\ref{video}(c). This push-pull mechanism is repeated for every cycle of the laser with a strong enough electric field in the density gradient direction, and is optimal when the gradient length is on the order of $\lambda_0/10$~\cite{thev16}.

\begin{figure}
\includegraphics[width=\columnwidth]{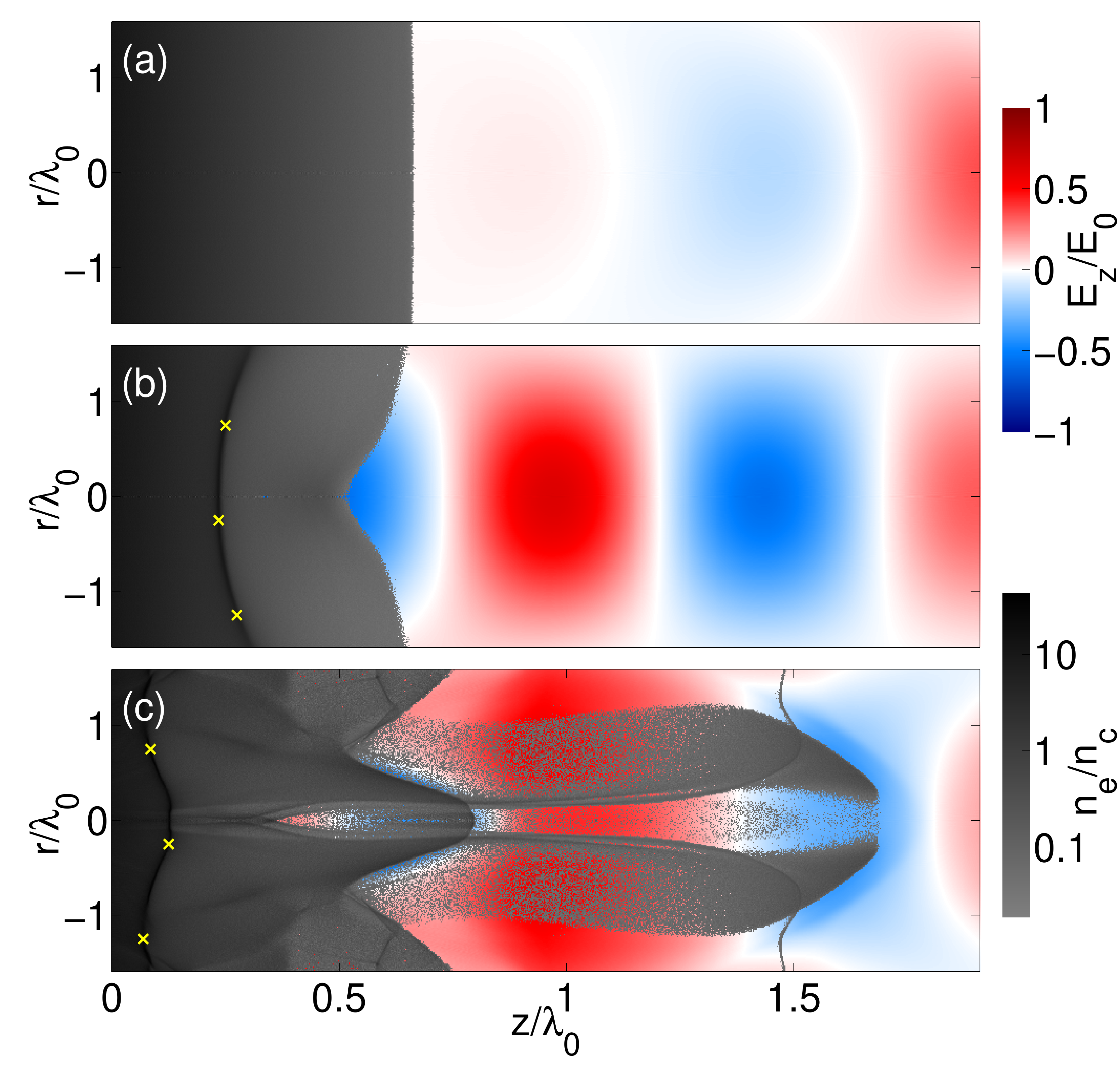}
\caption{PIC simulations showing the longitudinal electric field and electron density extracted at three different timesteps. (a) The plasma is still unperturbed by the incoming laser. (b) The electrons are pushed into the plasma by the laser, resulting in a density peak indicated by the yellows markers. (c) Electrons that were in the density peak are now pulled away from the plasma, between $z=1.2\lambda_0$ and $z=1.7\lambda_0$. (See also Supplementary Movie for more insight on the electron ejection)
}
\label{video}
\end{figure}

\begin{figure}
\includegraphics[width=\columnwidth]{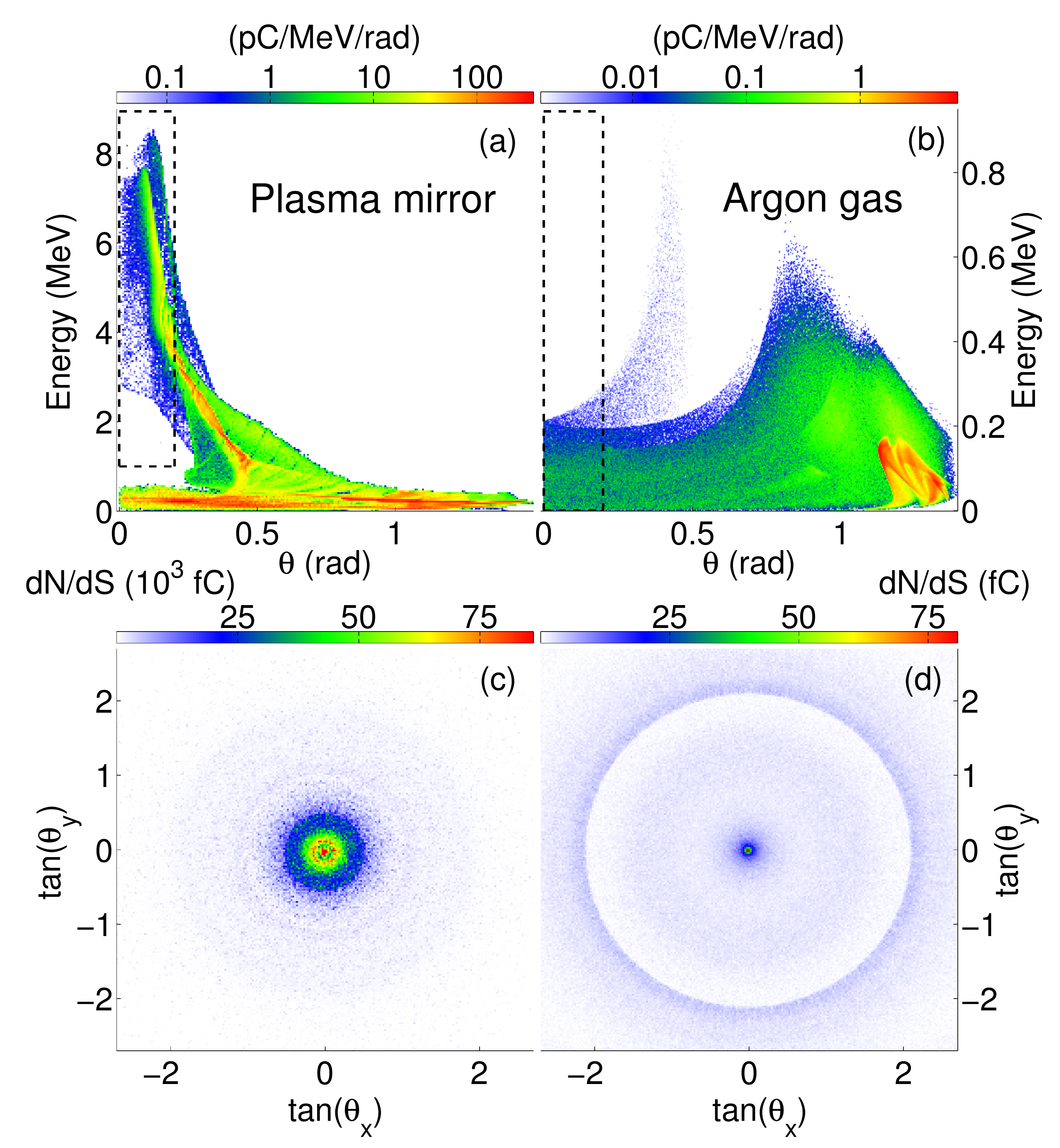}
\caption{(a)-(b) Energy-angle distribution and (c)-(d) angular distributions of the obtained electron beam after interaction of the pulse with (a),(c) a plasma mirror or (b),(d) argon gas. On the top images, the electrons inside the black rectangles are the one represented in Fig.~\ref{timecomp}. On the bottom images, the angular distribution is represented in the form $dN/dS$, with $dS = d\tan\theta_x \times d\tan\theta_y = d(p_x/p_z) \times d(p_y/p_z)$.}
\label{comparison}
\end{figure}

Figures~\ref{comparison}(a) and~\ref{comparison}(c) show the energy-angle and the angular distributions of the electrons ejected from the plasma mirror. The total ejected charge is about 60 pC. Furthermore, because they are injected into the laser with optimal initial conditions, a group of electrons representing several pC is accelerated to relativistic energies, typically from 1 to 8 MeV. At such relativistic speeds, the magnetic force $v_z \times B_\theta$ opposes the radial force $E_r$~\cite{marc13b}, mitigating the collimating effect of the radial polarization. These highly energetic electrons consequently form a ring shaped beam with a typical angle of 200 mrad with respect to the optical axis. This divergence angle is nonetheless significantly lower than what is achieved experimentally with linearly polarized lasers~\cite{thev15,bocoum16}. Analysis of the work done by the different forces shows that these electrons are accelerated by VLA in the reflected pulse by the longitudinal component of the electric field $E_z$, taking full advantage of the radial polarization. See Supplemental Material for more details.

\begin{figure}
\includegraphics[width=\columnwidth]{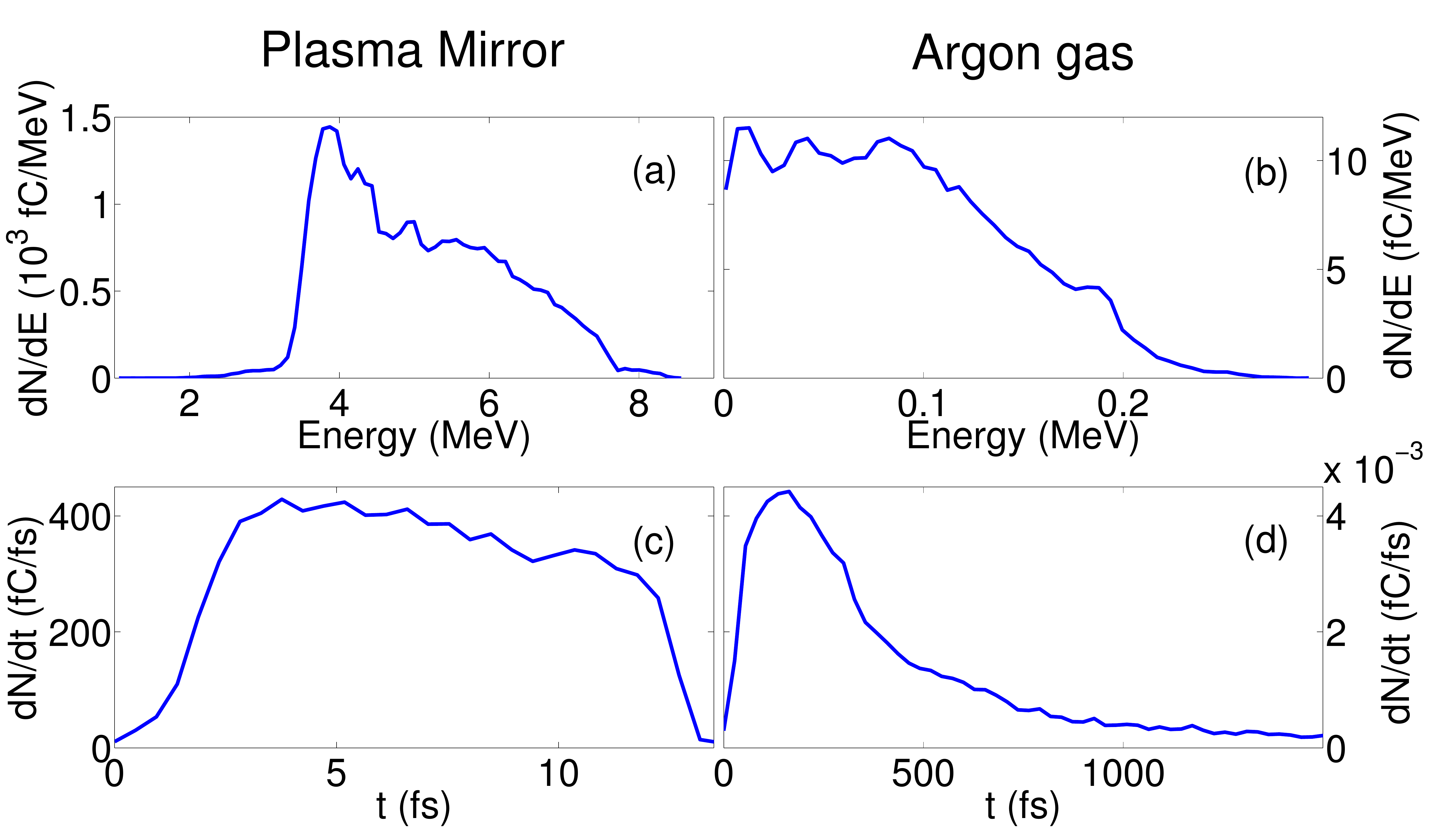}
\caption{(a)-(b) Energy spectrum and (c)-(d) time distribution of a chosen subset of electrons after $145 \mic$ of propagation for (a),(c) the plasma mirror simulation and (b),(d) the argon gas simulation.}
\label{timecomp}
\end{figure}

Figures~\ref{timecomp}(a) and~\ref{timecomp}(c) show the energy spectrum and time distribution of the electrons with $\theta < 200 \mrad$ and $E > 1 \mev$, after $145 \mic$ of propagation. Such filtering can typically be achieved using a pinhole and a magnet to select only certain angles and energies. These electrons represent 3.3 pC. Thanks to its high energy, this fast electron bunch is kept ultrashort, with a duration of around 12 fs. It is possible to diminish the duration of the pulse by reducing the acceptance angle of the selected electrons, at the cost of also diminishing the selected charge. To obtain the time distribution plots, electrons leaving the simulation are assumed to travel with constant speed afterwards.

In order to demonstrate the benefits of using a plasma mirror, we also perform PIC simulations of the ionization scheme. The same laser parameters are used, except for the absolute phase $\phi_0$, which is set to $\pi$ so as to have the same value as in~\cite{marc13}. The gas target is either hydrogen or argon. It is infinite in the transverse direction and is $10 \mic$ long in the longitudinal direction. The maximum electron density is chosen to be $3\times 10^{16}  \cc$, a value for which space-charge and plasma effects are negligible~\cite{marc13}.

In the case of hydrogen, we did not observe significant on-axis electron acceleration. This is because hydrogen has a low ionization energy, resulting in the electrons being ionized early in front of the laser pulse. This leads to a case very similar to that of Fig.~\ref{singlep}(a) and to a low final energy. In~\cite{marc13}, it was reported that electrons could be accelerated by ionizing a hydrogen target, but a much tighter focusing, $w_0 = 785 \nm$, that is harder to obtain in practice, was used in the simulations, resulting in a higher value for $a_{0z}$.

With argon however, some deeper shells electrons are generated inside the laser pulse, making it possible to accelerate on-axis electrons. Figures~\ref{comparison}(b) and~\ref{comparison}(d) show the energy-angle and the angular distributions of the obtained electrons. The total ejected charge is about 70 fC, three orders of magnitude lower than with the plasma mirror, and a few fC stay near the axis. The charge can nonetheless be increased by raising the initial gas density, at the cost of decreasing the electron beam quality~\cite{marc13}. The on-axis electrons are rather well focused and have an energy ranging from 0 to 200 keV, one to two orders of magnitude lower than with the plasma mirror. The energy spectrum and time distribution of the electrons with $\theta < 0.2 \rad$ after $145 \mic$ of propagation is shown in Figs.~\ref{timecomp}(b) and~\ref{timecomp}(d). This corresponds to a selected charge of 1.7 fC. As can be expected from its low energy and high energy spread, the time duration of this electron bunch is already over 300 fs after $145 \mic$ of propagation.

To conclude, plasma mirrors can inject high-charge electron beams with optimal conditions in the reflected pulse, making it possible to study experimentally VLA with radially polarized beams. Femtosecond, pC bunches of electrons with an energy between 3 and 7 MeV could be readily obtained with current kHz lasers. Moreover, we can take advantage of the high ejected charge to improve the quality of the electron beam, characterized by its transverse normalized emittance $\varepsilon_{n,x}$, by selecting only a subset of the ejected electrons. 
Depending on the application, a compromise between the charge, the energy spread and the normalized emittance of the beam can be found by filtering the electron beam, which could be achieved experimentally with an appropriate transport beam line. For instance, selecting electrons with an energy between 4.06 MeV and 4.14 MeV results in a $2\%$ energy spread, a charge of 100 fC and a normalized emittance of $\varepsilon_{n,x}  = \varepsilon_{n,y} = 0.18 \micron$. Emittance in the nm range can even be obtained by further filtering the beam while still maintaining the charge at the fC level. Such a low emittance ultrashort beam could be of a great interest for ultrafast electron diffraction experiments.


This work was funded by the European Research Council under Contract No. 306708, ERC Starting Grant FEMTOELEC and by the Agence Nationale pour la Recherche (ANR-14-CE32-0011-03 APERO). We acknowledge the support of CINES for access on super computer Occigen.

\bibliographystyle{apsrev}

\end{document}